%% file: main.tex
\documentclass{article} 

\usepackage[final]{colm2026_conference}

\usepackage{microtype}
\usepackage{hyperref}
\usepackage{url}
\usepackage{booktabs}
\usepackage{graphicx}
\usepackage{amsmath}
\usepackage{amssymb}
\usepackage{lineno}
\usepackage{mathrsfs}
\usepackage{bm}

\definecolor{darkblue}{rgb}{0, 0, 0.5}
\hypersetup{
    colorlinks=true,
    citecolor=darkblue,
    linkcolor=darkblue,
    urlcolor=darkblue
}

\title{Field-Aware Agent Skill Retrieval}

\author{Paimon Goulart$^{1}$ \quad Liang Wu$^{2}$ \quad Kelly Wan$^{2}$ \quad Evangelos E. Papalexakis$^{1}$ \\
\bfseries Liangjie Hong$^{2}$ \\[3pt]
\normalfont\small $^{1}$University of California, Riverside \quad $^{2}$Nokia \\
\normalfont\small \texttt{\{pgoul002,epapalex\}@ucr.edu} \quad \texttt{\{liang.wu,kelly.wan,liangjie.hong\}@nokia.com}
}

\begin{document}

\ifcolmsubmission
\linenumbers
\fi

\maketitle
\lhead{Paper published at the Lifelong Agent Workshop at COLM 2026}

\begin{abstract}
As lifelong learning agents accumulate lifelong growing skill banks, retrieving the correct skill becomes an increasingly important bottleneck. Most current skill retrieval methods treat each skill as one flat document by concatenating fields such as the name, description, and body. However, skills are naturally structured, multi-field objects, where each field provides different information about when and how the skill should be used. In this work, we study whether preserving this structure improves skill retrieval. We represent each skill as its separate components, and compute sparse and dense similarities for each field independently, exposing a naturally tensorized, field-aware representation of the skill bank. We then combine these field-level scores either with uniform weights or with a small learned MLP. Across two different skill retrieval benchmarks, SkillRet and SRA-Bench, we find that keeping fields separate improves hybrid retrieval, and learning over the field-level scores gives the strongest and most consistent results. Our field-aware MLP reaches $77.95$ Recall@10 on SkillRet and $83.78$ Recall@10 on SRA-Bench, outperforming the corresponding concatenated learned baselines. We also find that the advantage grows as the skill bank becomes larger, suggesting that field-aware skill retrieval becomes especially useful in the setting where retrieval is most difficult. Our results show that skill representation itself matters, and that simply preserving the structure already present in skill files can substantially improve retrieval.
\end{abstract}

\input{sec/introduction}
\input{sec/methods}
\input{sec/results}
\input{sec/conclusion}
\input{sec/acknowledgments}

\bibliography{refs}
\bibliographystyle{colm2026_conference}

\appendix

\end{document}

%% file: sec/introduction.tex
\section{Introduction}

Lifelong learning agents have lifelong growing skill banks. As an agent operates over time, it accumulates new skills, where each skill is a reusable and self-contained set of instructions that describes how to carry out certain tasks, typically stored as a \texttt{SKILL.md} file. Skills allow agents to continuously learn without requiring additional model retraining. However, a skill is only useful if the agent can actually find it. Every time the agent takes on a task, it has to search its skill bank and retrieve the correct skill(s) relevant to the task it needs to solve. At the same time, this step is becoming more automated and hidden from the user. Rather than having the user explicitly download and choose the skills they want, agentic systems are now creating, retrieving, and deploying skills/tools on their own whenever one is relevant to the current query or task. This leads to self-growing and evolving skill banks, which have been shown to improve model performance on specific tasks, especially when similar tasks are repeated later on \citep{yang2026autoskillexperiencedrivenlifelonglearning, wang2023voyageropenendedembodiedagent, lin2026museautoskillselfevolvingagentsskill, zhao2024expelllmagentsexperiential}.

This makes skill retrieval a potential bottleneck of current agentic frameworks. Since the agent now has to decide which skills to use, an error in retrieval could appear directly in the agent's behavior (i.e. an irrelevant skill can inject plausible but misleading instructions that the agent then strictly follows). This problem becomes much more apparent as the skill bank grows in size. As more skills accumulate, the likelihood of an agent selecting the wrong skill grows. This phenomenon has been referred to as skill shadowing \citep{song2026skillsworseagentsskill}, where a similar looking skill crowds out the one actually needed and degrades downstream performance. Here, retrieval quality determines whether a growing skill bank leads to better agent memory or a source of interference \citep{shi2025retrievalmodelsarenttoolsavvy}.

Despite how important this retrieval step is becoming, current skill retrieval methods continue to treat skills as regular flat documents. A skill is typically represented as a concatenated string of its fields, such as name, description, and body. From here, a lexical retriever, dense embedding method, or hybrid of the two is then applied in order to perform retrieval \citep{yang2026autoskillexperiencedrivenlifelonglearning, lin2026museautoskillselfevolvingagentsskill}. This ignores the fact that skills are naturally multi-component objects, which we believe is a huge oversight. The name and description often provide compact details for when/why a skill should be used, while the body contains detailed instructions, examples, and workflow steps. Flattening these fields removes this structure and dilutes the information contained within these fields.

In this work, we study whether preserving component structure improves skill retrieval. Inspired by methods such as BM25F and multi-field adaptive retrieval (mFAR), we represent each skill as three separate components and compute sparse and dense similarities for each component independently \citep{péreziglesias2009integratingprobabilisticmodelsbm25bm25f, li2025multifield, tang2026multifieldtoolretrieval}. This exposes a tensorized structure over the skill bank, keeping skills, features, and components separate rather than flattened. We find that this simple change improves retrieval over flattened baselines, and that a learned weighting over the component-level scores yields the strongest gains as the skill bank grows.

%% file: sec/methods.tex
\section{Methods}

\subsection{Tensorized skill representation}

Let the skill bank be $S=\{s_1,\dots,s_N\}$. Instead of treating each skill as one flat document, we keep its components separate. In our case, each skill has $C=3$ components: name, description, and body. For a component $c$ of skill $s_i$, we write its text as $x_{i,c}$.

For a given encoder $\phi$, each component is encoded independently,
\[
    e_{i,c}=\phi(x_{i,c}) \in \mathbb{R}^{d}.
\]
Stacking these representations over all skills and components gives a third-order skill tensor,
\[
    \boldsymbol{\mathscr{X}} \in \mathbb{R}^{N \times d \times C},
    \qquad
    \boldsymbol{\mathscr{X}}_{i,:,c}=e_{i,c}.
\]
The key difference from standard concatenation is that the component mode is preserved. A flattened retriever first collapses name, description, and body into one string and then encodes it, which would create a matrix. We instead keep these fields separate, so the retriever can still tell whether a match came from the name, the description, or the body, introducing a third mode.

Formalizing this skill bank as a tensor also gives us a useful structure to build on top of. Since the skill, feature, and component axes are kept separate, we can later use tensor methods to exploit relationships across these modes rather than only comparing skills as independent flat vectors. Previously, it has been shown that tensor decompositions can be useful for retrieval when the data naturally has multiple modes~\citep{chang-etal-2013-multi}. In our setting, this means that the same representation used for component-level retrieval could also be decomposed with methods such as CPD or Tucker. This can give us a compressed skill bank, faster retrieval, and a way to learn lower-dimensional structure across skills and components. Right now, we focus on showing that simply preserving the component mode already improves skill retrieval.

\subsection{Retrieval through tensor operations}

Given a query $q$, we encode it using the same encoder,
\[
    v_q = \phi(q) \in \mathbb{R}^{d}.
\]
We then compare the query to each component of each skill by taking a dot product along the feature dimension:
\[
    \mathbf{M}_{i,c}
    =
    \boldsymbol{\mathscr{X}}_{i,:,c}^{\top} v_q,
    \qquad
    \mathbf{M} \in \mathbb{R}^{N \times C}.
\]
After $\ell_2$ normalization, $\mathbf{M}_{i,c}$ is just the cosine similarity between the query and component $c$ of skill $s_i$. This gives one score per skill component.

To get a final relevance score for each skill, we combine over the component mode:
\[
    r_i
    =
    \sum_{c=1}^{C} w_c \mathbf{M}_{i,c}
    =
    (\mathbf{M}\mathbf{w})_i.
\]
With uniform weights, $w_c=1/C$, this is a training-free model that averages the component-level similarities. When supervision is available, we learn how to combine the component scores. In the simplest case this is a learned weight vector $\mathbf{w}$; in our strongest model, it is a small MLP,
\[
    r_i = f_\theta(\mathbf{M}_{i,:}).
\]
In both cases, the underlying skill representation is the same. The only difference is how the component-level similarities are combined.

\subsection{Sparse and dense views}

In practice, since the majority of skill retrieval pipelines use a form of hybrid retrieval, we use this same tensorized retrieval procedure for both sparse and dense representations. The sparse view uses TF-IDF features, while the dense view uses Qwen embeddings. Specifically we use Qwen$3$-Embedding-$0.6$B \citep{qwen3embedding}. Each view gives one component-score vector per skill:
\[
    \mathbf{M}^{\text{tfidf}}_{i,:}
    \quad \text{and} \quad
    \mathbf{M}^{\text{qwen}}_{i,:}.
\]
For a three-component skill, this gives six scores per query skill pair:
\[
    z_i =
    [
    \mathbf{M}^{\text{tfidf}}_{i,1},
    \mathbf{M}^{\text{tfidf}}_{i,2},
    \mathbf{M}^{\text{tfidf}}_{i,3},
    \mathbf{M}^{\text{qwen}}_{i,1},
    \mathbf{M}^{\text{qwen}}_{i,2},
    \mathbf{M}^{\text{qwen}}_{i,3}
    ].
\]
We then either average these scores uniformly or pass them through the same learned fusion head:
\[
    r_i = f_\theta(z_i).
\]
This lets the model combine hybrid sparse and dense evidence while still preserving which component each score came from.

%% file: sec/results.tex
\section{Results}

\subsection{Dataset}

For the evaluation of our method, we use two skill-retrieval benchmarks: SkillRet and SRA-Bench \citep{cho2026skillretlargescalebenchmarkskill, su2026skillretrievalaugmentationagentic}. SkillRet contains $6660$ candidate skills and $4997$ test queries, and also has a dedicated training set. SRA-Bench combines six individual datasets with domain specific skills: TheoremQA, LogicBench, ToolQA, MedCalcBench, CHAMP, and BigCodeBench. Combined this dataset has a total of $26262$ skills and $5400$ queries. For SRA-Bench we use a $70/30$ split per dataset and repeat these over five seeds and show the macro-average over all the datasets. For both datasets we report Hit@1, Recall@5, Recall@10, nDCG@5, nDCG@10, and MRR.

Following our method, each skill is split across $3$ components being name, description, and body. We encode each using TF-IDF and Qwen$3$-Embedding-$0.6$B, then combine the field-level similarities. We compare
three main families of methods. First, we compare standard retrieval baselines,
including BM25, BM25F (BM25F uses uniform field weights and is included as
a standard lexical field baseline), and concatenated TF-IDF, Qwen, and hybrid
retrieval. Second, we evaluate keeping fields
separate and averaging their similarities. Third, we evaluate a learned field-aware
model, which trains a small MLP over the field-level sparse and dense scores.

\begin{table}[t]
\centering\small
\caption{Retrieval on SkillRet (test, $4997$ queries). Best results are
shown in bold; second-best results are underlined. + MLP learns the
weighting of different scores.}
\label{tab:skillret}
\begin{tabular}{l cccccc}
\toprule
Method & Hit@1 & R@5 & R@10 & N@5 & N@10 & MRR \\
\midrule
BM25                       & 44.85 & 47.43 & 53.40 & 43.79 & 45.99 & 0.53 \\
BM25F                      & 46.83 & 48.61 & 54.55 & 45.22 & 47.42 & 0.55 \\
\midrule
TF-IDF (concat)            & 60.86 & 64.03 & 70.93 & 59.80 & 62.44 & 0.69 \\
TF-IDF (per-field)         & 63.60 & 65.74 & 73.75 & 61.88 & 64.93 & 0.71 \\
Qwen (concat)              & 53.39 & 55.55 & 61.88 & 52.33 & 54.68 & 0.62 \\
Qwen (per-field)           & 51.01 & 54.57 & 61.86 & 50.67 & 53.37 & 0.61 \\
Hybrid (concat)            & 66.98 & 67.08 & 72.82 & \underline{64.15} & 66.34 & \underline{0.75} \\
Hybrid (per-field)         & \underline{67.10} & \underline{67.29} & \underline{74.03} & 64.14 & \underline{66.72} & 0.74 \\
\midrule
Hybrid (concat) + MLP      & 64.44 & 66.58 & 73.61 & 62.79 & 65.48 & 0.73 \\
\textbf{Hybrid (per-field) + MLP} & \textbf{69.66} & \textbf{70.92} & \textbf{77.95} & \textbf{67.51} & \textbf{70.23} & \textbf{0.77} \\
\bottomrule
\end{tabular}
\end{table}

\begin{table}[t]
\centering\small
\caption{Retrieval on SRA-Bench, macro-averaged over the six datasets
(per-dataset $70/30$, five seeds). Best results are shown in bold;
second-best results are underlined. + MLP learns the weighting of different scores.}
\label{tab:sra}
\begin{tabular}{l cccccc}
\toprule
Method & Hit@1 & R@5 & R@10 & N@5 & N@10 & MRR \\
\midrule
BM25                       & 38.30 & 50.77 & 59.79 & 43.92 & 47.14 & 0.47 \\
BM25F                      & 39.21 & 51.09 & 60.05 & 44.48 & 47.70 & 0.48 \\
\midrule
TF-IDF (concat)            & 39.07 & 55.37 & 64.75 & 47.20 & 50.55 & 0.49 \\
TF-IDF (per-field)         & 33.29 & 47.58 & 55.62 & 40.46 & 43.41 & 0.43 \\
Qwen (concat)              & 45.25 & 52.35 & 58.77 & 48.46 & 50.84 & 0.52 \\
Qwen (per-field)           & 47.33 & 60.12 & 69.00 & 53.55 & 56.84 & 0.56 \\
Hybrid (concat)            & 52.84 & 63.48 & 70.56 & 57.76 & 60.41 & 0.61 \\
Hybrid (per-field)         & 50.76 & 65.72 & 74.49 & 57.64 & 60.86 & 0.60 \\
\midrule
Hybrid (concat) + MLP      & \underline{55.04} & \underline{68.32} & \underline{76.52} & \underline{61.42} & \underline{64.42} & \underline{0.64} \\
\textbf{Hybrid (per-field) + MLP} & \textbf{55.90} & \textbf{74.60} & \textbf{83.78} & \textbf{65.35} & \textbf{68.80} & \textbf{0.67} \\
\bottomrule
\end{tabular}
\end{table}

\subsection{Analysis}

As shown in Tables~\ref{tab:skillret} and \ref{tab:sra}, keeping the fields separate yields useful gains, especially once sparse and dense embeddings are combined. On SkillRet, the training-free hybrid improves the concatenated hybrid on both Recall@10 ($74.03$ vs.\ $72.82$) and nDCG@10 ($66.72$ vs.
$66.34$). On SRA-Bench, we again see an increase in Recall@10 from $70.56$ to
$74.49$ and nDCG@10 from $60.41$ to $60.86$.

Additionally, we compare learning the weighting over flattened scores and per-field scores. On SkillRet, we can see that learning over the concatenated hybrid scores does not improve over the concatenated hybrid version. However, once the model is able to weight per-field scores, the learned version gives the best result across every metric. The Hybrid (per-field) + MLP model reaches $69.66$ Hit@1, $77.95$ Recall@10, $70.23$ nDCG@10, and $0.77$ MRR. Similarly on SRA-Bench, the per-field MLP shows a large overall increase in performance. Recall@10 increases from $76.52$ to $83.78$, and nDCG@10 increases from $64.42$ to $68.80$.

These results support the idea that skill retrieval benefits when the components of a skill are kept separate. The gains are small, but become much larger once we learn how to properly weight the different fields of a skill.

\subsection{Scaling results}

We also want to evaluate how our proposed method behaves as the skill bank grows in size. For each query, we keep the correct skill in the skill bank and add randomly sampled distractor skills until reaching the desired skill bank size. This allows us to evaluate whether a method still retrieves the correct skill when the skill bank becomes larger and more noisy.

\begin{figure}[t] \centering \includegraphics[width=0.49\textwidth]{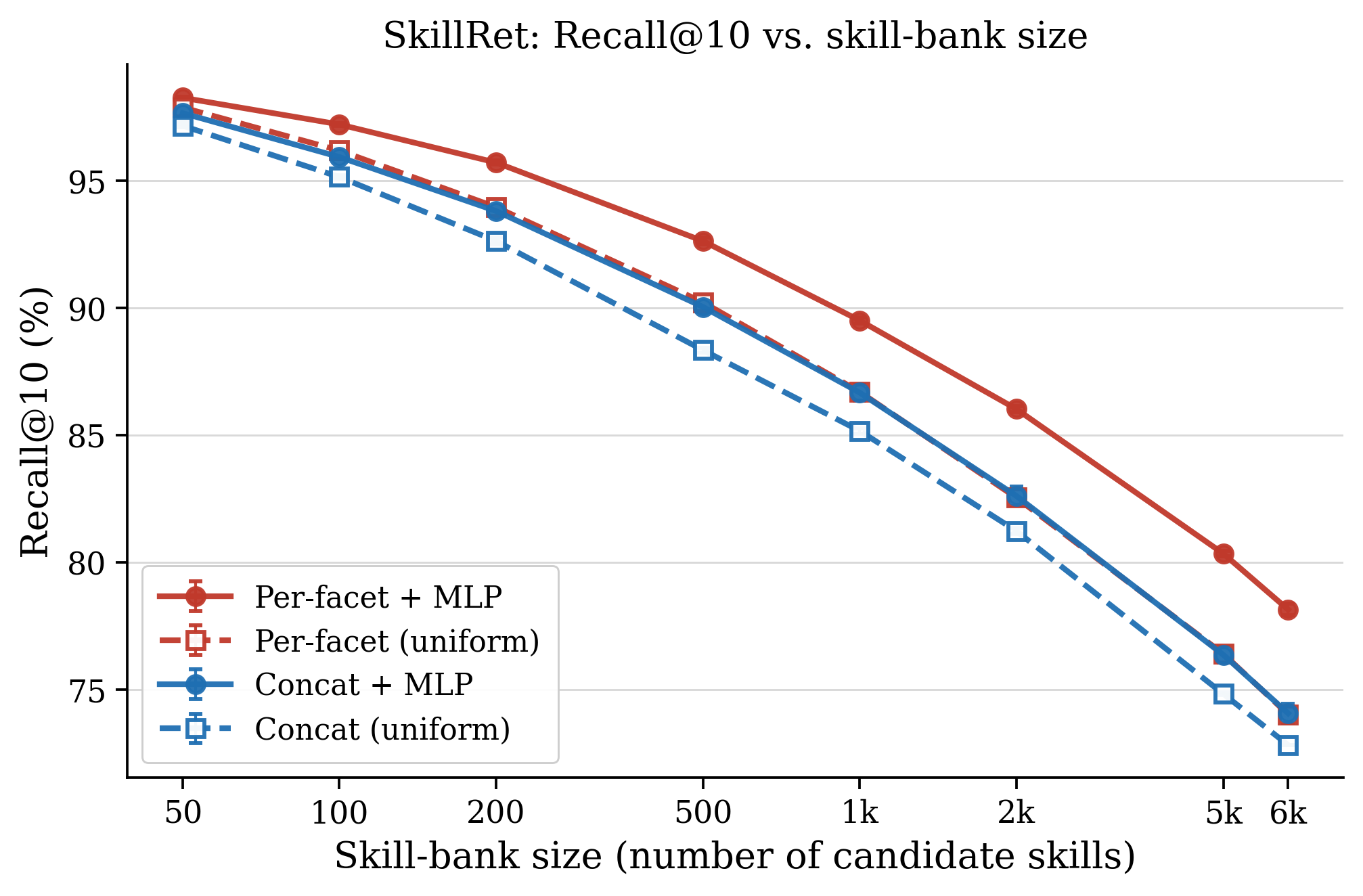} \includegraphics[width=0.49\textwidth]{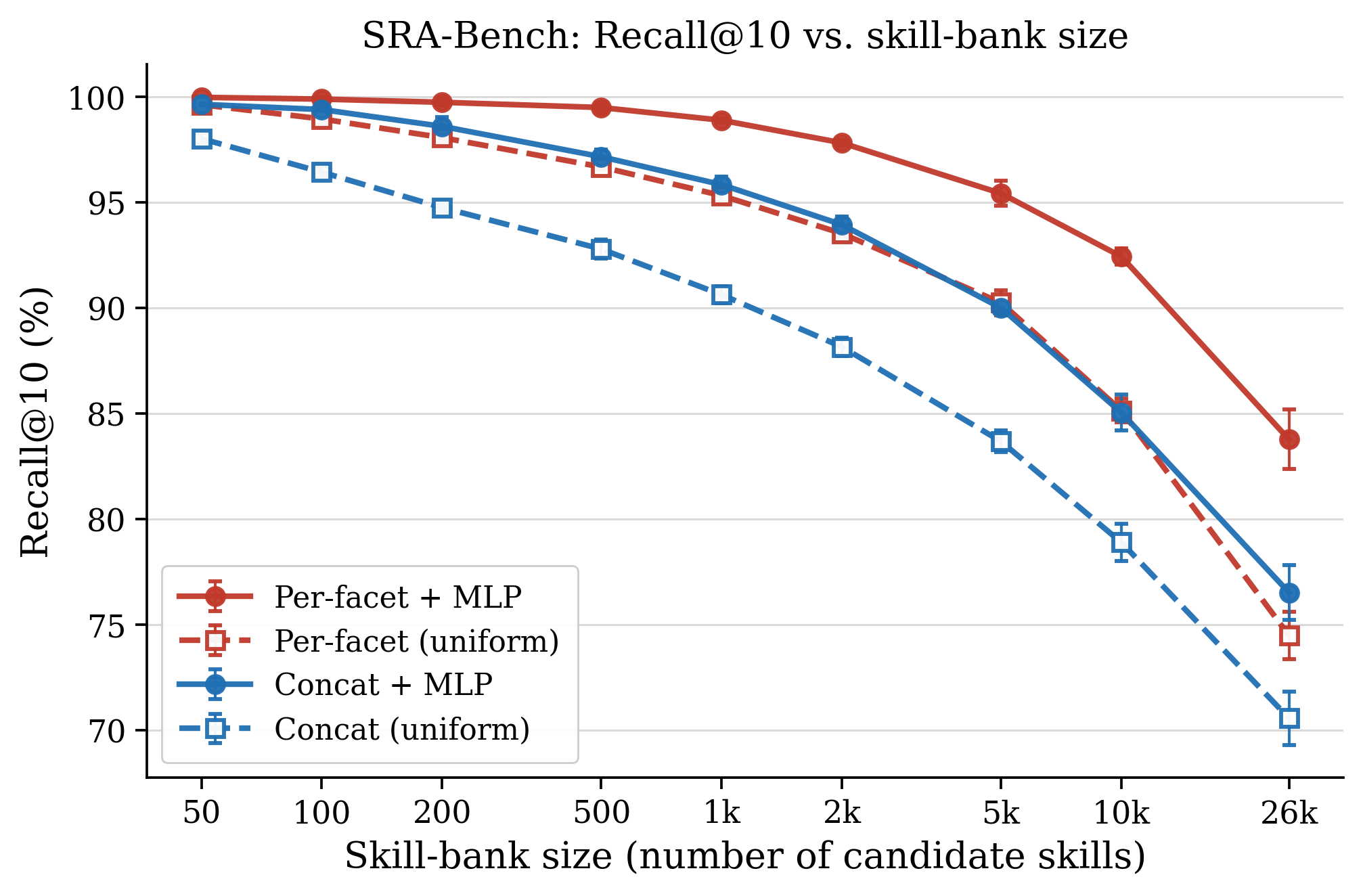} \caption{Recall@10 as the skill-bank size increases on SkillRet (left) and SRA-Bench (right). The per-field MLP stays strongest as the bank grows, with the largest gap appearing at the full-bank setting.} \label{fig:scaling} \end{figure}

Since we are interested in retrieval, we decided to focus on Recall@10. Figure~\ref{fig:scaling} shows Recall@10 as the skill bank increases in size. As we expect, all methods decrease in performance as more skills are added. As the skill bank grows, more similar and partially related skills can crowd out the correct one. However, the per-field MLP is the most stable across both datasets. On SkillRet, the field-aware MLP stays above the other methods across the whole curve. At the full SkillRet bank size, it remains around $78\%$ Recall@10, while the concatenated MLP and the uniform methods are lower. The field-aware uniform weighted model also remains close to the concatenated MLP for pretty much the entire curve, which again suggests that preserving field structure alone is already useful. On SRA-Bench we see even clearer results. At small bank sizes, all the methods perform well; however, the gap becomes larger as the skill bank grows in size. At $26262$ skills, the per-field MLP reaches $83.78$ Recall@10, compared to $76.52$ for the concatenated MLP, $74.49$ for the per-field uniform model, and $70.56$ for the concatenated uniform model. We consistently see that field-aware retrieval matters more as the skill bank grows in size.

%% file: sec/conclusion.tex
\section{Conclusion}

In this work, we study skill retrieval from the perspective of skill representation, which we believe is a current limitation. Instead of combining each skill into one flat document, we preserve the structure that already exists in skills by splitting each skill into its individual components: name, description, and body. Across SkillRet and SRA-Bench, we find that this simple change improves retrieval over concatenated approaches. The training-free per-field hybrid already gives gains, and learning a small MLP to properly weight the per-field scores gives the strongest results. These gains become even more important as the skill bank grows, suggesting that field-aware retrieval is especially useful as retrieval becomes more difficult.

Due to the impacts of this per-field approach, we take this as initial evidence that skill representation really does matter for retrieval. Even by simply splitting skills by field and learning a small model over these fields leads to large gains. While many skill retrieval methods focus on concatenating skill fields or fine-tuning embedding models on these concatenated representations, our results demonstrate that this can create a bottleneck and misrepresentation problem. Future work will explore how to further exploit this per-field representation, either by learning a new embedding model that is field-aware, or by using the tensorized structure of the skill bank more directly.

%% file: sec/acknowledgments.tex
\section*{Acknowledgments}
Research at the University of California, Riverside was supported in part by the National Science Foundation under CAREER grant no. IIS 2046086 and grant no. 2524228.